# Mid-infrared polarized emission from black phosphorus light-emitting diodes

*Junjia Wang[1], Adrien Rousseau[1], Mei Yang[1], Tony Low[2], Sébastien Francoeur[1], and Stéphane Kéna-Cohen[1]*

[1]Department of Engineering Physics, Polytechnique Montréal, Montréal, Québec, H3C 3A7, Canada

[2]Department of Electrical and Computer Engineering, University of Minnesota, Minneapolis, MN, 55455, USA

The mid-infrared (MIR) spectral range is of immense use for civilian and military applications. The large number of vibrational absorption bands in this range can be used for gas sensing, process control and spectroscopy. In addition, there exists transparency windows in the atmosphere such as that between 3.6–3.8 μm, which are ideal for free-space optical communication, range finding and thermal imaging. A number of different semiconductor platforms have been used for MIR light-emission. This includes InAsSb/InAs quantum wells[1], InSb/AlInSb[2], GaInAsSbP pentanary alloys[3], and intersubband transitions in group III-V compounds[4]. These approaches, however, are costly and lack the potential for integration on silicon and silicon-on-insulator platforms. In this respect, two-dimensional (2D) materials are particularly attractive due to the ease with which they can be heterointegrated. Weak interactions between neighbouring atomic layers in these materials allows for deposition on arbitrary substrates and van der Waals heterostructures enable the design of devices with targeted optoelectronic properties. In this Letter, we demonstrate a light-emitting diode (LED) based on the 2D semiconductor black phosphorus (BP). The device, which is composed of a BP/molybdenum disulfide ($MoS_2$) heterojunction emits polarized light at $\lambda$ = 3.68 μm with room-temperature internal and external quantum efficiencies (IQE and EQE) of ~1% and ~$3\times10^{-2}$ %, respectively. The ability to tune the bandgap, and consequently emission wavelength of BP, with layer number, strain and electric field make it a particularly attractive platform for MIR emission.

Electroluminescence (EL) from 2D transition metal dichalcogenides (TMDs) was observed shortly after monolayers from this class of materials were first isolated[5, 6, 7, 8, 9]. In monolayer TMD crystals, the formation of a direct bandgap allows for reasonable light-emission efficiencies to be achieved. The high degree of confinement in monolayers also ensures a large exciton binding energy and consequently efficient excitonic emission from the fabricated devices. Monolayer LEDs have been fabricated using a broad range of geometries: Schottky junctions,[5] p-n junctions defined using local gates[7, 8, 10] and ionic-liquid gating[11]. However, electroluminescence from planar devices typically leads to low EQEs of <0.1%. Vertical p-n junctions based on atomically flat heterojunctions can be fabricated using van der Waals stacking techniques.[6] This leads to simpler device geometries and higher efficiencies. It also allows for unique architectures to be fabricated such as those based on hBN tunnel barriers[12]. In the latter, single monolayer devices have led to EQEs of ~1%, where as multiple quantum well geometries can achieve EQEs of up to ~8.4%.[12]

Black phosphorus has recently attracted significant attention as an anisotropic material for optoelectronic and electronic applications[13]. This semiconducting material shows relatively high carrier mobilities and a direct bandgap regardless of film thickness[13, 14, 15, 16, 17]. The atomic crystal structure of BP consists of layered $sp^3$-hybridized phosphorus atoms with distinct configurations along two orientations termed the armchair (AC) and zigzag (ZZ) directions. This gives rise to strong in-plane anisotropy for thermal conduction[18, 19], carrier transport and optical absorption[20, 21, 22, 23, 24]. In the context of optoelectronics, one of the most interesting properties of BP is its widely tunable bandgap and the fact that it is a direct bandgap material from the bulk down to bilayers. Thick multilayer BP flakes have a direct band gap ~0.3 eV[15, 25] and when reducing the number of layers, this gap increases to nearly ~2 eV.[26] In addition, electrostatic doping or electric fields can be used to further reduce the bandgap from its bulk value.[27, 28, 29] Until now, much of the focus on BP optoelectronics has focused on its use as a material for MIR photodetection. In this work, we demonstrate that BP also holds great promise for electrically-driven MIR light emission.

Our device architecture is shown schematically in Figure 1a and b. It consists of 70 nm-thick p-type BP layer on a 10 nm-thick n-type $MoS_2$ layer fabricated using the hot pick-up and stamping technique[30] (see Methods). Both layers are contacted with Cr/Au electrodes. The device was fabricated on a highly p-doped Si substrate with 90 nm $SiO_2$ oxide layer, which can be used as a back gate. An optical micrograph of the fabricated LED is shown in Fig. 1c with both layer outlined. Fig. 2a shows the equilibrium band diagram of the $BP/MoS_2$ heterostructure assuming Anderson's rule for band alignment and an electron affinity for BP of $\chi \sim 3.6$ eV.[31] The *p* and *n*-doping of BP and $MoS_2$, respectively, leads to a band alignment favoring the interface accumulation of holes in BP and electrons in $MoS_2$. As previously observed for thick BP flakes, two distinct regimes of operation are possible.[32] For negative $V_{DS}$, band-to-band tunneling can occur, as shown in Fig. 2b. When $V_{DS}$ is made more negative, the BP valence band minimum shifts above the $MoS_2$ conduction band minimum. An energy window emerges where electrons from the BP valence band can transit directly to the $MoS_2$ conduction band. As shown in Fig. 2c, for positive $V_{DS}$, the current is mostly due to thermionic crossing of the barrier by electrons from $MoS_2$ to BP. Figures 2d and 2e show the resulting I-V characteristic of the device measured under ambient conditions. Note that the forward bias behavior can be strongly modulated by applying a back gate voltage, which varies the doping density. Figure 2d highlights that the on-off ratio can be as high as $10^5$.

The photoluminescence spectrum of BP, shown in Fig. 3a (top), was measured using above gap excitation at $\lambda = 808$ nm with an irradiance 100 $\mu W/\mu m^2$. The emission intensity maximum is at 3.68 $\mu m$, which agrees well with reported measurements on thick BP flakes[33]. To measure EL against the thermal background, a $V_{GS} = 10$ $V_{pk-pk}$ sinusoidal signal is applied to modulate $I_{DS}$ and consequently the EL intensity. The EL is measured using FTIR with a HgCdTe detector and standard lock-in techniques. Figure 3a (bottom) shows the EL spectrum measured at a forward bias $V_{DS}$ = 7V. The EL spectrum is centered at the same wavelength as the PL. The spectral linewidth appears narrower that that in PL, but this is most-likely related to the periodic noise minima at the EL spectrum edges, which are due to the interferometer scan length. Figure 3b (top) shows the IV characteristic of the device measured when the gate is modulated. As in Fig. 2d, the forward bias current saturates at large $V_{DS}$. Figure 3b (bottom) shows the LED radiance, which increases concomitantly with the increasing current for positive $V_{DS}$. By measuring the absolute emitted power and correcting for losses through the reflective measurement objective, we find an external quantum efficiency (EQE) of 0.03% at $V_{DS}$ = 7 V (see Supplementary for details). Figure S7 shows the EQE as a function of the forward bias current. A small increase in EQE is observed when the carrier density is increased, which is consistent with competition between monomolecular and bimolecular recombination.

In addition, selection rules near the bandgap of BP lead to a much larger oscillator strength for polarization along the AC direction. In EL, we similarly find that the luminescence is indeed polarized mostly along the AC direction. Figure 3d shows the measured intensity as a function of polarization. We find a polarization ratio of ~3 for the AC to ZZ intensity.

A large fraction of the photon loss can be ascribed to the optical environment of the LED. When materials with different dielectric constants are in the near field of an emitting dipole, emission will be directed into each material in proportion to its optical density of states, which scales as $n^3$. For our structure, emission will preferentially be radiated in the bottom direction, due to the higher refractive index of silicon compared to air. Figure 3c shows the calculated radiation pattern for a horizontal BP dipole in our structure (along the AC direction).[34] We calculate an outcoupling efficiency of only 3.6%, which corresponds to an IQE of ~1%.

In summary, we have demonstrated a room-temperature BP/MoS$_2$ LED emitting at 3.68 μm, which corresponds to a strategically important MIR spectral region. The device shows strongly polarized emission along the AC crystal direction. Several strategies can be used to increase the efficiency of this structure. Notably, the outcoupling efficiency would be significant improved via the use of a low-index (e.g. glass) or metallic substrate and the IQE could be further improved via the use of a double heterostructure or carrier-selective contacts. The ability to heterointegrate BP on various platforms, and to tune the bandgap of BP, *e.g.* via the BP thickness, a vertical electrical bias or chemical doping, make it an extremely attractive option for use as a versatile MIR light source.

METHODS

**Device fabrication**

BP (Black Phosphorus, Smart Element) was mechanically exfoliated onto a substrate using a PVC tape (SPV 224PR-M, Nitto Denko) inside a glovebox, and the flake is picked up by a stamp consisting of a layer of polycarbonate (PC) mounted on a block of polydimethylsiloxane (PDMS) at a temperature of 60 degrees. The $MoS_2$ was then picked up using the BP flake at a temperature of 80 degerees. The heterostructure is released onto a commercial p-type silicon substrate (Silicon/Silicon dioxide (90 nm) wafers from Graphene Supermarket) at a temperature of 180 degrees followed by removal of the PC using cholorform. Then the electrodes patterns are defined by electron beam lithography (EBL, Raith eLine) using PMMA A4 495 resist. 10 nm of chromium and 60 nm of gold were thermally evaporated to form the contacts (EvoVac, Angstrom Engineering). Following lift-off, 1 nm of aluminium was deposited on top of the structure to form a $Al_2O_3$ oxide oxidation barrier.

The thickness of the flakes were measured by atomic force microscopy (AFM, Bruker Dimension FastScan). An optical microscope image of the completed device is shown in Fig. 1c. The BP is on top of $MoS_2$ and the thickness are 70 nm and 10 nm, respectively. The active area is ~285.2 $um^2$. Raman spectroscopy was used to determine the axis of the flake (see Supplementary for details).

**Device characterization**

A schematic of the experimental setup is shown in Fig. S1. Completed devices were mounted onto a translation stage and contacted by electrical probes. Current–voltage measurements were taken in a two-probe configuration using a source meter (2614B, Keithley). For the PL measurements, the sample was excited by focusing light beam from a 808 nm semiconductor laser. A microscope objective lens was used to focus the beam and a chopper is placed before to modulate the light at a frequency of 500 Hz. The PL was then reflected by a $CaF_2$ beam splitter and collected by the MCT detector. The AC output signal from the MCT detector was connected to a lock-in amplifier (MFLI, Zurich Instruments) and the DC output form the lock-in amplifier was connected back to the MCT detector and the FTIR. The lock-in amplifier used a time constant of 3 seconds with an amplification factor of 1000 and the FTIR step-scan used a delay of 24 seconds and averaged 20 times. For the EL measurements, the two electrodes were contacted by the two probes and the back of the chip was connected by a sinusoidal AC signal (10 $V_{pk\text{-}pk}$) sychronised to the lock-in amplifer with a frequency of 500 Hz. The EL is collected by the microscope objective lens and reflected by the $CaF_2$ beam splitter to the MCT detector. The lock-in amplifier and FTIR step-scan used same settings as the PL measurements except the amplification factor was 2000.

AUTHOR INFORMATION

**Corresponding Author**

*s.kena-cohen@polymtl.ca

**Notes**

The authors declare no competing financial interest.

ACKNOWLEDGMENT

The authors would like to acknowledge M.Z.U Khan for providing the marked substrates. S.K.C. acknowledges support from the Canada Research Chairs program. S.K.C.and S.F. gratefully acknowledge funding for this work from NSERC Strategic Grant STPGP-506808.

**FIGURES**

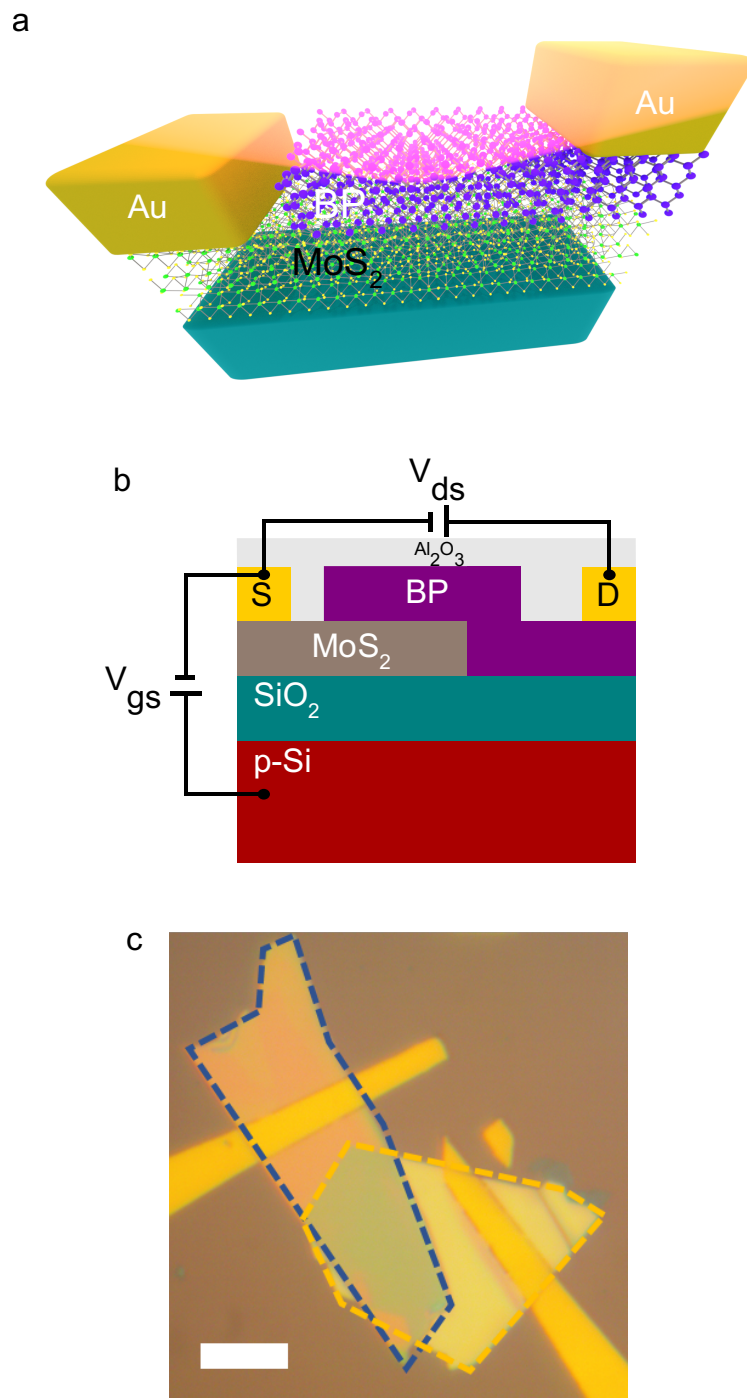

Figure 1. **BP/MoS$_2$ p-n junctions.** (**a**) A schematic illustration of BP/MoS$_2$ heterojunction p-n diode. (**b**) A schematic illustration the cross-sectional view of the BP/MoS2 heterojunction device. (**c**) Optical image of the BP/MoS$_2$ heterojunction. BP: flake with blue outline. MoS$_2$ flake with orange outline. Scale bar: 10 μm.

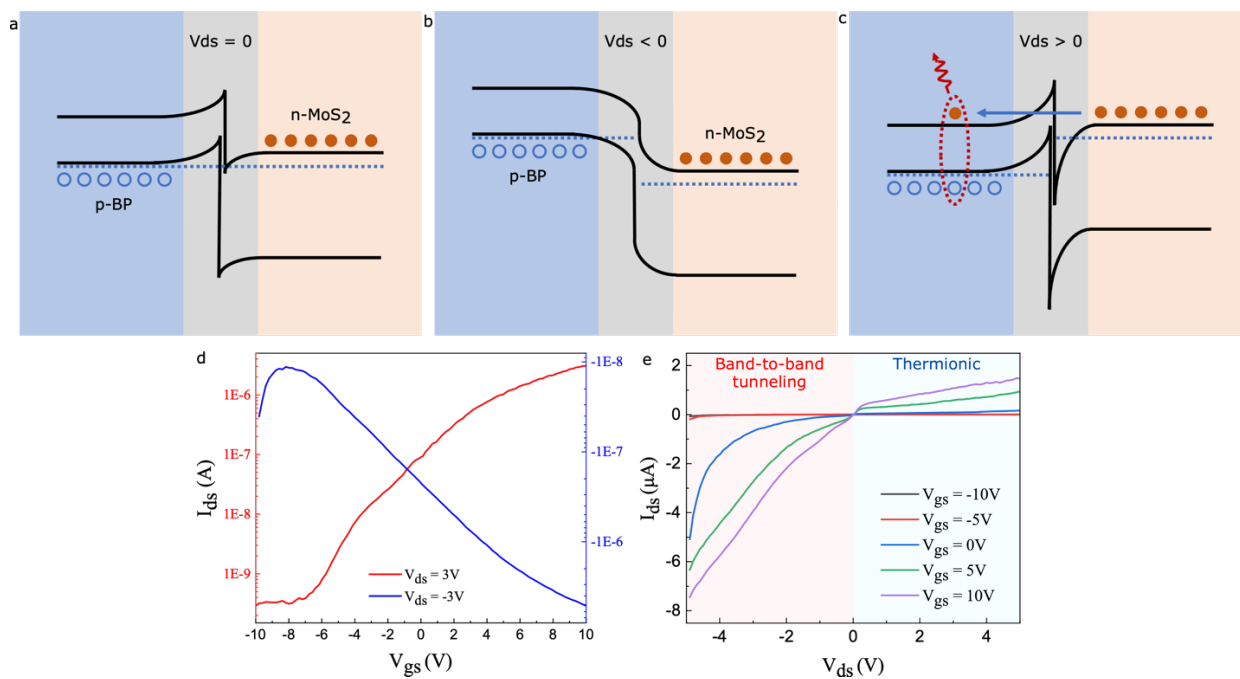

Figure 2. **Electrical characterization**. (**a**) Equilibrium band diagram of the BP/MoS$_2$ heterojunction assuming the electron affinity rule. (**b**) Band diagram under negative VDS showing the possibility of band-to-band tunneling. (**c**) Band diagram at positive VDS showing the light-emission process. (**d**) Transistor transfer characteristic for positive and negative drain-source voltage. (**e**) Transistor output characteristic for varying gate voltage.

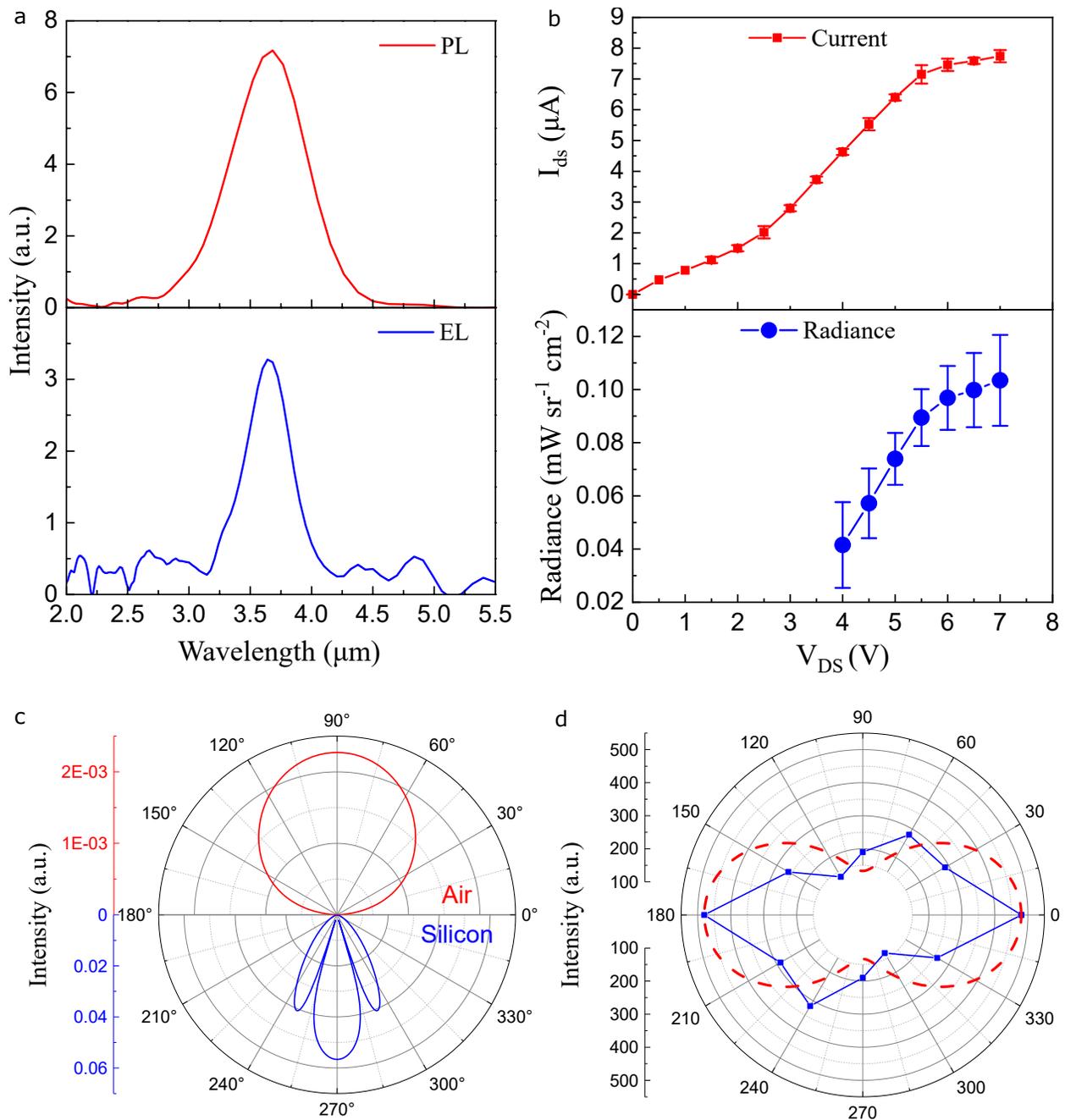

Figure 3. (**a**) Photoluminiscience (red) and electroluminiscience (blue) of the BP/MoS$_2$ diode. (**b**) Upper: forward bias current Lower: current under different forward bias voltages measured with lock-in detection at the gate modulation frequency. (**c**) Simulated outcoupling efficiency for a horizontal dipole emitting in the top (air) and bottom (silicon) directions. (**d**) Polarization-resolved measurements of the EL intensity.